\documentclass[superscriptaddress, aps,prb,amsmath,twocolumn,amssymb,titlepage]{revtex4-1}
\usepackage{graphicx}
\begin{document}

\title{Random number generation using spontaneous symmetry breaking in a Kerr resonator}

\author{Liam Quinn}
\affiliation{Physics Department, The University of Auckland, Auckland 1142, New Zealand}
\affiliation{The Dodd-Walls Centre for Photonic and Quantum Technologies}
\author{Gang Xu}
\affiliation{Physics Department, The University of Auckland, Auckland 1142, New Zealand}
\affiliation{The Dodd-Walls Centre for Photonic and Quantum Technologies}
\author{Yiqing Xu}
\affiliation{Physics Department, The University of Auckland, Auckland 1142, New Zealand}
\affiliation{The Dodd-Walls Centre for Photonic and Quantum Technologies}
\author{Zongda Li}
\affiliation{Physics Department, The University of Auckland, Auckland 1142, New Zealand}
\affiliation{The Dodd-Walls Centre for Photonic and Quantum Technologies}
\author{Julien Fatome}
\affiliation{Laboratoire Interdisciplinaire Carnot de Bourgogne, UMR 6303 CNRS-Université de Bourgogne, Dijon, France}
\author{Stuart G. Murdoch}
\affiliation{Physics Department, The University of Auckland, Auckland 1142, New Zealand}
\affiliation{The Dodd-Walls Centre for Photonic and Quantum Technologies}
\author{St\'{e}phane Coen}
\affiliation{Physics Department, The University of Auckland, Auckland 1142, New Zealand}
\affiliation{The Dodd-Walls Centre for Photonic and Quantum Technologies}
\author{Miro Erkintalo}
\affiliation{Physics Department, The University of Auckland, Auckland 1142, New Zealand}
\affiliation{The Dodd-Walls Centre for Photonic and Quantum Technologies}

%\affiliation[1]{Physics Department, The University of Auckland, Auckland 1142, New Zealand}
%\affiliation[2]{The Dodd-Walls Centre for Photonic and Quantum Technologies}
%\affiliation[3]{Laboratoire Interdisciplinaire Carnot de Bourgogne, UMR 6303 CNRS-Université de Bourgogne, Dijon, France}

%\affiliation[*]{Corresponding author: liam.quinn@auckland.ac.nz}

\begin{abstract}
We experimentally demonstrate an all-optical random number generator based on spontaneous symmetry breaking in a coherently-driven Kerr resonator. Random bit sequences are generated by repeatedly tuning a control parameter across a symmetry-breaking bifurcation that enacts random selection between two possible steady-states of the system. Experiments are performed in a fibre ring resonator, where the two symmetry-broken steady-states are associated with orthogonal polarization modes. Detrimental biases due to system asymmetries are completely suppressed by leveraging a  recently-discovered self-symmetrization phenomenon that ensures the symmetry breaking acts as an unbiased coin toss, with a genuinely random selection between the two available steady-states. We optically generate bits at a rate of over 3~MHz without post-processing and verify their randomness using the National Institute of Standards and Technology and Dieharder statistical test suites.
\end{abstract}

%\setboolean{displaycopyright}{true}

\maketitle

\renewcommand{\baselinestretch}{0.9} 

Coherently-driven nonlinear resonators have attracted significant attention over the last decade due to their potential applications, including the generation of optical frequency combs \cite{delhayeOpticalFrequencyComb2007, kippenbergMicroresonatorBasedOpticalFrequency2011, sternBatteryoperatedIntegratedFrequency2018}, as well as the rich physics and dynamics  they exhibit, involving phenomena such as dissipative solitons and complex phase transitions \cite{coenTemporalCavitySolitons2015, garbinDissipativePolarizationDomain2021, yuSpontaneousPulseFormation2021}. In this context, one universal nonlinear phenomenon that has stimulated particular recent interest is spontaneous symmetry breaking (SSB), with pioneering experimental observations reported in monolithic microresonators \cite{delbinoSymmetryBreakingCounterpropagating2017,caoExperimentalDemonstrationSpontaneous2017, woodleyUniversalSymmetrybreakingDynamics2018}, macroscopic fibre ring resonators \cite{garbinAsymmetricBalanceSymmetry2020, xuSpontaneousSymmetryBreaking2021}, and coupled-cavities \cite{bgarbinSpontaneousSymmetryBreaking2022}. SSB refers to the ubiquitous process whereby a physical system in a symmetric state loses its symmetry in favour of two asymmetric states -- a phenomenon that can be accessed by tuning a control parameter across a bifurcation point. In the context of systems involving a single resonator \cite{delbinoSymmetryBreakingCounterpropagating2017,caoExperimentalDemonstrationSpontaneous2017, woodleyUniversalSymmetrybreakingDynamics2018, garbinAsymmetricBalanceSymmetry2020, xuSpontaneousSymmetryBreaking2021}, SSB has been observed between two counter-propagating fields \cite{delbinoSymmetryBreakingCounterpropagating2017, caoExperimentalDemonstrationSpontaneous2017, woodleyUniversalSymmetrybreakingDynamics2018}, as well as between co-propagating orthogonal polarization modes \cite{xuSpontaneousSymmetryBreaking2021, garbinAsymmetricBalanceSymmetry2020}, with the detuning between the driving laser and a cavity resonance (or the power of that laser) acting as the control parameter \cite{caoExperimentalDemonstrationSpontaneous2017, garbinAsymmetricBalanceSymmetry2020}. In each case, slight power deviations between the two competing fields circulating in the cavity are magnified over consecutive roundtrips, leading to the selection of one of the possible asymmetric states \cite{haeltermanPolarizationMultistabilityInstability1994, delbinoSymmetryBreakingCounterpropagating2017}. Under perfectly symmetric operating conditions, the selection enacted by SSB is completely random, alluding to the possibility of using the phenomenon for the generation of random numbers that are critical for various applications, including cryptography and security \cite{ stojanovskiChaosbasedRandomNumber2001,  yuanRandomnessRequirementClauserHorneShimonyHolt2015}. Unfortunately, in practice, this prospect is typically prohibited as a result of the sensitivity of SSB to residual asymmetries, which leads to a statistical preference towards one of the states, thus preventing truly random selection upon symmetry-breaking \cite{crauelAdditiveNoiseDestroys1998, garbinAsymmetricBalanceSymmetry2020}.

Remarkably, a phenomenon was recently discovered that can completely eliminate the impact of asymmetries on polarization SSB dynamics in fibre ring resonators. Specifically, it was shown in \cite{coenNonlinearTopologicalSymmetry2023} that the implementation of a $\pi -$phase shift between two orthogonal polarization modes of the resonator resulted in a period-2 (P2) switching behaviour between two polarization modes in the cavity, somewhat similar to the period doubling regime used in bulk optical parametric oscillators \cite{ steinmeyerObservationPerioddoublingSequence1994, steinleUnbiasedAllOpticalRandomNumber2017}. This periodic switching leads to a natural 'self-symmetrization' of the system, as asymmetries are averaged out over two roundtrips. While the results in \cite{coenNonlinearTopologicalSymmetry2023} presented convincing evidence of the randomness of the SSB process and highlighted the potential for random number generation, the system was not yet optimised for fast random number generation and lacked a thorough analysis of the statistical properties of the generated bit sequence.

Here we report on a comprehensive experimental study of random number generation based on SSB in a coherently-driven Kerr resonator under conditions of self-symmetrized operation. Building upon the experiment reported in \cite{coenNonlinearTopologicalSymmetry2023}, we construct an optimised setup that allows us to generate random optical bits at rates exceeding 3 MHz without post-processing by repeatedly scanning the cavity detuning across the SSB bifurcation point. We judiciously demonstrate the randomness of the system by showing that the generated sequences pass all pertinent tests from the National Institute of Standards and Technology and Dieharder Statistical Test Suites \cite{rukhinStatisticalTestSuite2001, brownDieharder2018}. Our scheme offers future potential for random number generation with GHz repetition rates and requires no post-processing. Moreover, thanks to its all-optical nature, our scheme could enable random number generation with improved speed and energy efficiency compared to current state-of-the-art systems \cite{marandiAllopticalQuantumRandom2012, okawachiQuantumRandomNumber2016, steinleUnbiasedAllOpticalRandomNumber2017}. In addition, our scheme is an all-fiber system operating at 1550~nm, which makes it an excellent candidate for integration with other optical devices. By demonstrating the true randomness of the SSB process, our results also underline the potential of the self-symmetrized mechanism for the realization of a novel coherent Ising \cite{wangCoherentIsingMachine2013, inagakiCoherentIsingMachine2016, mcmahonFullyProgrammable100spin2016} or Potts machine \cite{honari-latifpourOpticalPottsMachine2020}. 

\begin{figure}[t]
	\centering
	\includegraphics[width=1.0\linewidth]{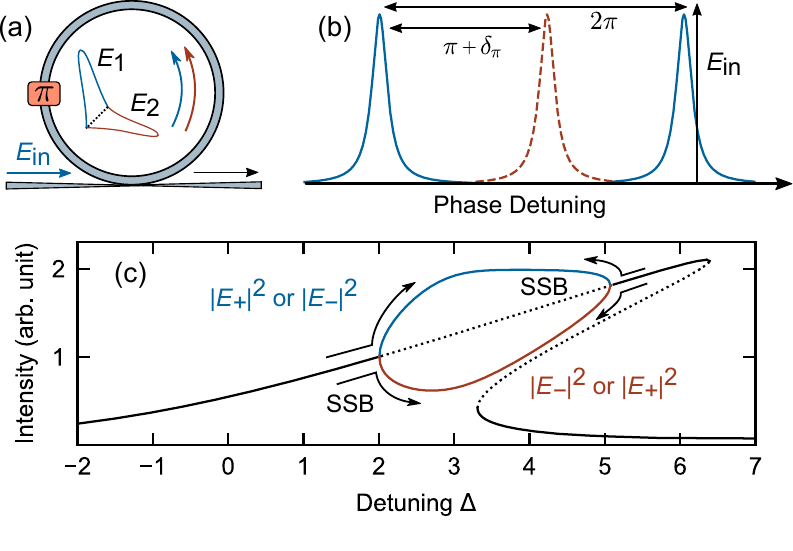}
	\caption{\textbf{(a)} Schematic illustration of a passive, coherently-driven Kerr resonator supporting two orthogonal polarization modes. Only one mode is driven, and a $\pi$ phase-defect separates the resonances of the two cavity modes $E_1$ and $E_2$. \textbf{(b)} Cavity resonances for the two principal polarization modes $E_1$ and $E_2$. $E_{\text{in}}$ corresponds to the driving field, while $\delta_\pi$ describes the deviation of the mode resonance separation from $\pi$. \textbf{(c)} Bifurcation diagram, showing the intensities of the intracavity steady-states in the circular basis. The solutions are symmetric $(|E_+|^2 = |E_-|^2)$ for small and large detunings, but become asymmetric via SSB. Solid (dashed) curves represent stable (unstable) solutions.}
	\label{figure_1}
\end{figure}

We first briefly summarise the physics of the self-symmetrized SSB scheme that underpins our experiments \cite{coenNonlinearTopologicalSymmetry2023}. To this end, Fig.~\ref{figure_1}(a) shows a conceptual illustration of our SSB-based random number generator (RNG). The system is built around a passive,  coherently-driven nonlinear Kerr resonator supporting two principal orthogonal polarization modes ($E_1$ and $E_2$) -- defined as polarization modes that map back to themselves at the end of each roundtrip. Only one of the principal modes is driven, and the cavity resonances corresponding to the two modes are offset by a phase-detuning of approximately $\pi$, realized via a birefringent defect (polarization controller) within the cavity. To understand the dynamics of the system, we consider the intracavity field evolution in the circular polarization basis, defining the left- and right-circular polarization components as $E_\pm = (E_1 \pm iE_2)/\sqrt{2}$. As a result of the $\pi$-phase shift between the two cavity modes, the sign of one of these components (e.g. $E_2$) becomes inverted relative to the other at each roundtrip $E_2 \longrightarrow E_2e^{i\pi} = -E_2$. This alternating effect leads to a periodic swapping of the circular polarization projections such that, at the end of each roundtrip, $E_+$ and $E_-$ swap values. Over two roundtrips, each circular polarization component returns to their original value, and it was shown in \cite{coenNonlinearTopologicalSymmetry2023} that the evolution of the components, when averaged over two roundtrips, obeys coupled Lugiato-Lefever -like mean-field equations, which are well-known to display SSB \cite{lugiatoSpatialDissipativeStructures1987}. Importantly the two-roundtrip periodicity of the nonlinear motion leads to self-symmetrization that eliminates all asymmetries, leading to the components $E_\pm$ experiencing identical detuning and driving terms. This complete self-symmetrization remains robust even in the presence of both an imperfect $\pi$-phase shift [referred to as $\delta_\pi$ in Fig.~\ref{figure_1}(b)] and an imperfectly aligned driving polarization \cite{coenNonlinearTopologicalSymmetry2023}.

Figure \ref{figure_1}(c) displays predicted steady-state intensity values of the $E_\pm$ components obtained from solely even or odd round-trips, with parameters similar to our experiments. Here, the intensities are plotted versus the detuning between the frequency of the driving laser and a closest cavity resonance of mode $E_1$, normalized to half the resonance linewidth. At low detunings, the two polarization projections $E_\pm$ are exactly equal. Symmetry breaking occurs above a certain threshold, manifesting itself through the parting of the two intensities of the polarization modes. Once the detuning is sufficiently large, the symmetry-broken 'bubble' closes, and the intensity levels become equal again. Thanks to the self-symmetrization that occurs as a result of the $\pi$-phase defect, repeated scanning of a control parameter [e.g. cavity detuning as in Fig. \ref{figure_1}(c)] leads to random selection between the two possible states in the $E_\pm$ basis in a way that is immune to asymmetries \cite{garbinAsymmetricBalanceSymmetry2020}. By measuring the intensity of one of the circular polarization projections $E_\pm$ and assigning a value of '1' to high intensity measurements and '0' to low intensity measurements, the mechanism can be used for robust generation of random bit sequences. Our all-optical RNG is founded on this principle. 

\begin{figure}[t]
	\centering
	\includegraphics[width=1.0\linewidth]{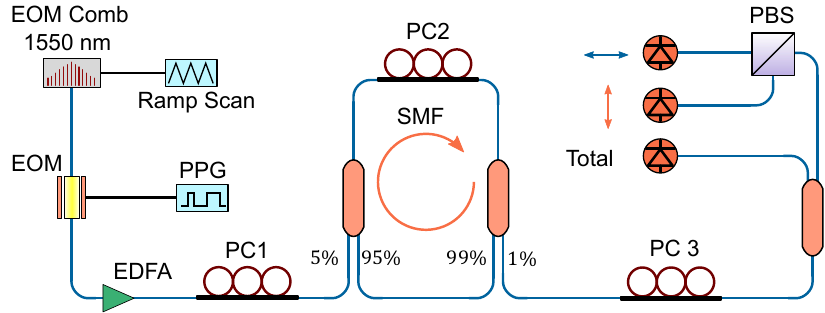}
	\caption{Experimental setup. PPG: pulse pattern generator, PC: polarization controller, EDFA: erbium doped fibre amplifier, PBS: polarization beam splitter, EOM: Electro-optic modulator, SMF: Single mode fibre.}
	\label{Schem}
\end{figure}

Figure \ref{Schem} illustrates the key components of our experimental setup. The cavity consists of 57~m of MetroCor single-mode fibre with a Kerr nonlinearity coefficient of $\gamma = 2.5~\text{W}^{-1}\text{km}^{-1}$. The fiber exhibits normal group-velocity dispersion at the driving wavelength of 1552~nm, thus ensuring suppression of modulation instabilities. A polarization controller [PC1 in Fig.~1] is used to align the driving field along one of the principal polarization modes of the cavity. A second, intra-cavity polarization controller [PC2 in Fig. 2] is used to introduce a phase defect of $\pi$ between the two principal polarization modes. A 95:5 coupler injects the input field into the cavity, and a 99:1 coupler extracts a part of the circulating field, yielding a total cavity finesse of 40. After the 99:1 coupler, we project the output field onto the $E_\pm$ basis using a polarization controller [PC3 in Fig.~\ref{Schem}], before separating these two components using a polarization beam splitter (PBS) and measuring their intensity profiles by means of fast-photodetectors connected to a real-time oscilloscope.

The resonator is coherently-driven with a train of 2-ps-long pulses derived from an electro-optic (EO) comb generator seeded with a narrow-linewidth continuous-wave (CW) laser at 1552~nm. The EO comb generator consist of a cascade of two phase and one amplitude modulator followed by a nonlinear pulse compression stage. The repetition rate of the EO comb is derived from an external RF clock synthesizer that is carefully adjusted to be a large integer multiple of the cavity free-spectral range (FSR) of $3.655~\text{MHz}$. We use an additional electro-optic modulator driven by a pulse-pattern generator to control the repetition rate of the driving pulses as desired before they are injected into the resonator. The experiments that will follow used an injection repetition rate of 1.17~GHz, such that there are 320 pulses circulating in the cavity simultaneously, separated in time by 0.8~ns. Each of the pulses undergoes SSB independently, thus allowing for an increased rate of random number generation via time-multiplexing. In this context, we emphasize that even though the pulses are only 2-ps-long, they undergo SSB in a manner that is qualitatively similar to the SSB of the CW states visualised in Fig. \ref{figure_1}(c).

The experiments that will follow used an injection repetition rate of 1.17~GHz, such that there are

In order to repeatedly scan the cavity detuning back and forth across the SSB bifurcation point, we first actively stabilize the detuning within the symmetry-breaking regime using the technique described in \cite{nielsenInvitedArticleEmission2018}. We subsequently use an acousto-optic modulator (AOM) to sinusoidally modulate the frequency of the main pump beam, resulting in periodic scanning across the SSB bifurcation point at a frequency of 10~kHz. This causes the system to periodically switch between symmetric and symmetry-broken regimes, with each pulse randomly selecting one of the two possible solution states beyond the SSB bifurcation point. By detecting the intracavity pulse intensities following the bifurcation, a sequence of random bits is obtained. The 10~kHz modulation frequency and 320 pulses per roundtrip result in a generation of random bits at a speed of 3.2~Mbit/s.

%Briefly, a low-power probe beam is extracted from the driving laser, frequency shifted with an acousto-optic modulator (AOM), and launched into the resonator such that it counter-rotates with respect to the main pump; the detuning is stabilized by locking the portion of the probe beam output from the resonator to a set value using a PID controller.

Although based upon the same principle as the setup used in \cite{coenNonlinearTopologicalSymmetry2023}, our current configuration has been optimised to enable random number generation at significantly higher data rates than in \cite{coenNonlinearTopologicalSymmetry2023}. Thanks to their much shorter duration, the electro-optically generated 2~ps pulses can be more closely spaced and more efficiently amplified, thus allowing for more efficient time-multiplexing and correspondingly higher data rates than the original experiment, which relied on 1.1-ns-long quasi-cw pulses. Our cavity is also five times longer than the one used in \cite{coenNonlinearTopologicalSymmetry2023}, which further increases the number of pulses that can simultaneously circulate in the resonator.

\begin{figure}[t]
	\centering
	\includegraphics[width=1.0\linewidth]{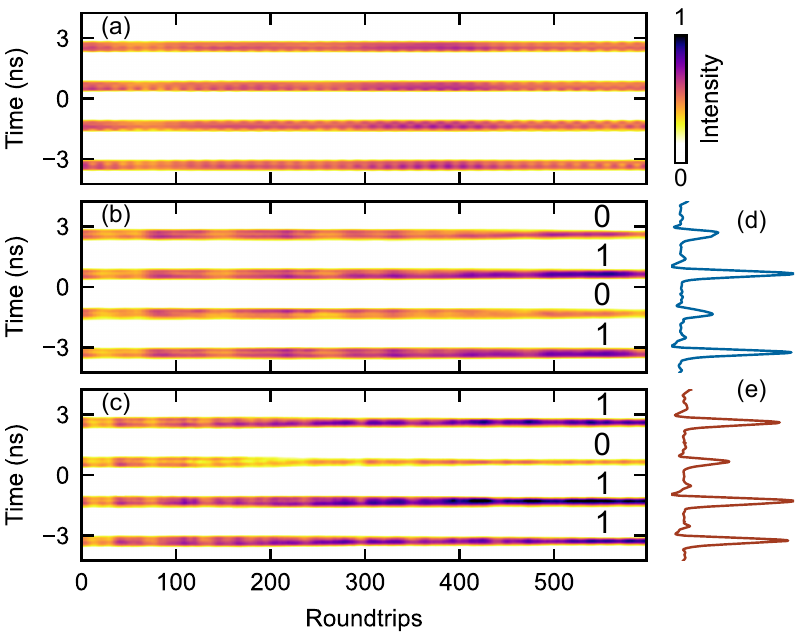}
	\caption{\textbf{(a)} Evolution of the intracavity temporal intensity profiles corresponding to \textit{one} of the circular polarization states as the detuning is scanned over consecutive roundtrips. \textbf{(b)-(c)} Evolution of the intracavity temporal intensity profiles when considering every second roundtrip for two independent realizations of the RNG system. \textbf{(d)-(e)} Temporal intensity profiles at the final roundtrip for each realization. }
	\label{fast_time}
\end{figure}

\begin{figure*}[t]{}
	\centering
	\includegraphics[width=\linewidth]{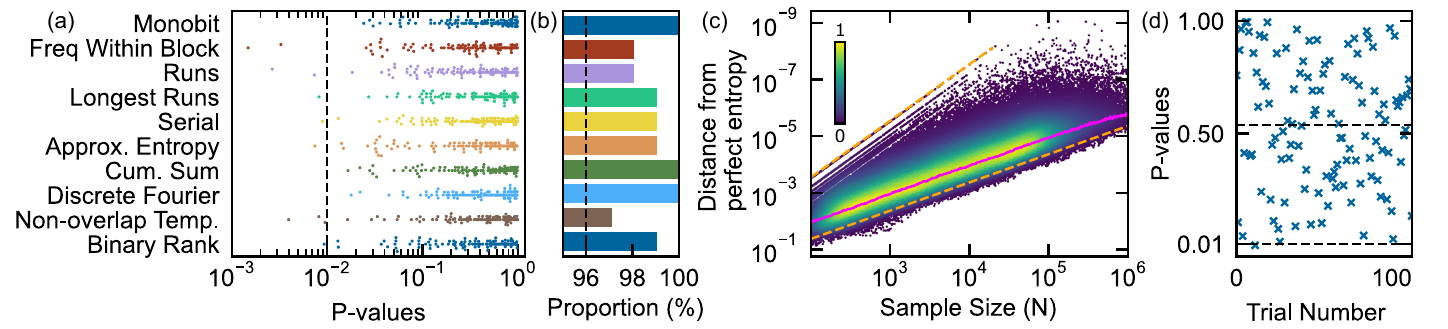}
	\caption{Test results for randomness. \textbf{(a)} P-values generated by applying the NIST statistical tests to 100 samples of a 32 million bit sequence.  \textbf{(b)} proportion of trials that pass the tests with the significance level of 0.01, which should be greater than 96$\%$.  \textbf{(c)} Shannon entropy in the generated bit-stream. The top and bottom orange lines show one bit-flip away from perfect entropy, and one standard deviation from the expected entropy generated by a random sequence, respectively. The individual points on the graph indicate the distance from perfect entropy for a given sample size, and the color gradient highlights the density of overlapping points. Finally, the magenta line shows the average of all individual trials. \textbf{(d)} P-values generated by application of the full suite of Dieharder Statistical tests. Horizontal dashed lines show the average of the test results (0.53) and the failure threshold (0.01).}
	\label{optimised}
\end{figure*}

Figure \ref{fast_time} shows illustrative experimental results that demonstrate the generation of random bit sequences in our system. In these experiments, we recorded a long time series in real time on an oscilloscope as the detuning was scanned across the SSB bifurcation point. We then sliced the time series into individual segments spanning a single roundtrip: the pseudocolour plot in Fig. \ref{fast_time}(a) corresponds to a horizontal concatenation of these segments to reveal the spatio-temporal evolution of the intracavity intensity of the circular polarization states (with only four bits shown for clarity). The SSB dynamics are hidden by the period-2 alternation that occurs in the measurement of a single circular polarization state, and is thus not directly visible in Fig. \ref{fast_time}(a); however, when we only consider every second roundtrip, we clearly see the emergence of two states with differing intensity [Fig. \ref{fast_time}(b,d)]. Figure \ref{fast_time}(c,e) shows SSB results from a second experimental realization, demonstrating the generation of a new, independent bit sequence.

%\begin{figure}[t]
%	\centering
%	\includegraphics[width=\linewidth]{fig3_shorter.pdf}
%	\caption{Test results for randomness using the NIST Statistical Test Suite \cite{rukhinStatisticalTestSuite2001}. \textbf{(a)} P-values for each individual test are shown for all 100 sub-sequences. A p-value of 0.01 or less (black dotted line) indicates a failure for the individual trial. \textbf{(b)} proportion of sequences that pass the tests with the significance level of 0.01. The black dotted line indicates the NIST standard recommendation of a minimum pass rate of 96$\%$. }
%	\label{optimised}
%\end{figure}

To confirm the true randomness of the bit sequences generated by our scheme, we implemented three separate statistical tests. First, Fig.~\ref{optimised}(a) shows results from the National Institute of Standards and Technology (NIST) Statistical test Suite for RNGs (NIST STS-2.1.2) \cite{rukhinStatisticalTestSuite2001} applied to a sample of 32 million bits. Each test is applied to 100 samples from the entire 32 million-bit sequence, and each of these trials returns a p-value --- the probability of this result occurring under the null hypothesis of a truly random source. NIST recommends that the minimum proportion of trials with a p-value greater than 0.01 be at least 96$\%$ for any given test. The bit sequences generated from our system pass this condition for every test~[Fig.~\ref{optimised}(b)].\\

As a second test, we consider the entropy generated by the system. In the ideal situation, each bit has an entropy of 1, which implies that each generated bit simulates a perfectly random coin toss. Following the method outlined by Steinle et al. \cite{steinleUnbiasedAllOpticalRandomNumber2017}, we show in Fig.~\ref{optimised}(c) the conditional entropy --- a measure of the system's memory --- computed for our sequence of 32 millions bits. Each data point shows the distance from perfect entropy for a single trial of sample size $N$, while the top and orange lines show bounds for one bit-flip from perfect entropy and one standard deviation from the expected conditional Shannon entropy, respectively. We calculate the entropy considering both the conditional probability of individual bits \textbf{1} and \textbf{0}, as well as the probabilities for the tuples \textbf{11, 10, 01, 00} (e.g. the probability of observing the tuple \textbf{11} given that it follows the tuple \textbf{00}). We find that both the mean (magenta line), and the vast majority of the individual trials exhibit entropy values within one standard deviation of what is expected for an unbiased system, which gives us confidence that each bit generated can be used as a random bit without further processing \cite{steinleUnbiasedAllOpticalRandomNumber2017}.

Finally, Fig.~\ref{optimised}(d) shows the results of the Dieharder tests applied to our bit stream \cite{brownDieharder2018}. Each Dieharder test consists of 100-1000 trials applied to subsets of our input data. A Kolmogorov-Smirnov (KS) test is then applied to the p-values obtained by each of these tests to determine the probability of observing the distribution of p-values in the random sampling of a uniform distribution. These KS p-values are plotted in Figure \ref{optimised}(d). The random distribution of p-values centred at 0.53 and only a single failure occurring at the 0.01 significance level over all 113 tests shows that we have no evidence against the null hypothesis that our RNG is a truly random source.

%Before closing, we discuss the limits of our system. To ensure genuine random number generation, it is important that the detuning modulation rate is significantly slower than the photon lifetime. This ensures that the system spends sufficiently long in the symmetric state to allow the self-symmetrization mechanism to erase any memory from the previous passage through the SSB point. Likewise, a sufficiently slow modulation rate is needed to ensure the SSB has sufficient time to fully develop prior to the detection of the intra-cavity intensities (in our experiments, we synchronize the measurement with the maxima of the modulation to ensure maximal time spent in the SSB region).

To conclude, we have experimentally demonstrated an all-optical random number generator based on polarization symmetry breaking in a coherently-driven passive Kerr resonator. We achieved bit generation rates in excess of 3 MHz without any post-processing, which we believe can be further increased by optimising the cavity design and the modulation used to scan across the SSB bifurcation. Preliminary analysis suggests that operation in a microresonator platform has the potential to enable bit generation rates in the GHz range. We verified the randomness of the generated bit sequences using several statistical tests. In addition to representing a new avenue for physical random number generation, our work suggests that polarization symmetry breaking could be used to realise novel optical Ising or Potts machines. \\

%\textbf{Funding.} Dodd-Walls Centre for Photonic and Quantum Technologies. Marsden Funding (18-UOA-310) from the Royal Society Te Apārangi. CNRS (IRP Wall-IN project), FEDER-FSE Bourgogne 2014/2020 programs, Conseil Régional de Bourgogne Franche-Comté.

\textbf{Acknowledgements.} 

We acknowledge the financial support provided by The Royal Society of New Zealand in the form of Marsden Funding (18-UOA-310). Additional financial contributions were kindly provided by CNRS through the IRP Wall-IN project, the FEDER-FSE Bourgogne 2014/2020 programs and the Conseil Régional de Bourgogne Franche-Comté. 

%\textbf{Disclosures.} The authors declare no conflicts of interest. 

%\textbf{Data Availability.} The underlying data in this paper is not publicly available at this time, however it may be obtained from the authors upon reasonable request.

%\renewcommand{\baselinestretch}{0.9} 
\bibliography{bibliography_randomness}

\begin{thebibliography}{29}%
\makeatletter
\providecommand \@ifxundefined [1]{%
 \@ifx{#1\undefined}
}%
\providecommand \@ifnum [1]{%
 \ifnum #1\expandafter \@firstoftwo
 \else \expandafter \@secondoftwo
 \fi
}%
\providecommand \@ifx [1]{%
 \ifx #1\expandafter \@firstoftwo
 \else \expandafter \@secondoftwo
 \fi
}%
\providecommand \natexlab [1]{#1}%
\providecommand \enquote  [1]{``#1''}%
\providecommand \bibnamefont  [1]{#1}%
\providecommand \bibfnamefont [1]{#1}%
\providecommand \citenamefont [1]{#1}%
\providecommand \href@noop [0]{\@secondoftwo}%
\providecommand \href [0]{\begingroup \@sanitize@url \@href}%
\providecommand \@href[1]{\@@startlink{#1}\@@href}%
\providecommand \@@href[1]{\endgroup#1\@@endlink}%
\providecommand \@sanitize@url [0]{\catcode `\\12\catcode `\$12\catcode
  `\&12\catcode `\#12\catcode `\^12\catcode `\_12\catcode `\%12\relax}%
\providecommand \@@startlink[1]{}%
\providecommand \@@endlink[0]{}%
\providecommand \url  [0]{\begingroup\@sanitize@url \@url }%
\providecommand \@url [1]{\endgroup\@href {#1}{\urlprefix }}%
\providecommand \urlprefix  [0]{URL }%
\providecommand \Eprint [0]{\href }%
\providecommand \doibase [0]{http://dx.doi.org/}%
\providecommand \selectlanguage [0]{\@gobble}%
\providecommand \bibinfo  [0]{\@secondoftwo}%
\providecommand \bibfield  [0]{\@secondoftwo}%
\providecommand \translation [1]{[#1]}%
\providecommand \BibitemOpen [0]{}%
\providecommand \bibitemStop [0]{}%
\providecommand \bibitemNoStop [0]{.\EOS\space}%
\providecommand \EOS [0]{\spacefactor3000\relax}%
\providecommand \BibitemShut  [1]{\csname bibitem#1\endcsname}%
\let\auto@bib@innerbib\@empty
%</preamble>
\bibitem [{\citenamefont {Del'Haye}\ \emph {et~al.}(2007)\citenamefont
  {Del'Haye}, \citenamefont {Schliesser}, \citenamefont {Arcizet},
  \citenamefont {Wilken}, \citenamefont {Holzwarth},\ and\ \citenamefont
  {Kippenberg}}]{delhayeOpticalFrequencyComb2007}%
  \BibitemOpen
  \bibfield  {author} {\bibinfo {author} {\bibfnamefont {P.}~\bibnamefont
  {Del'Haye}}, \bibinfo {author} {\bibfnamefont {A.}~\bibnamefont
  {Schliesser}}, \bibinfo {author} {\bibfnamefont {O.}~\bibnamefont {Arcizet}},
  \bibinfo {author} {\bibfnamefont {T.}~\bibnamefont {Wilken}}, \bibinfo
  {author} {\bibfnamefont {R.}~\bibnamefont {Holzwarth}}, \ and\ \bibinfo
  {author} {\bibfnamefont {T.~J.}\ \bibnamefont {Kippenberg}},\ }\href@noop {}
  {\bibfield  {journal} {\bibinfo  {journal} {Nature}\ }\textbf {\bibinfo
  {volume} {450}},\ \bibinfo {pages} {1214} (\bibinfo {year}
  {2007})}\BibitemShut {NoStop}%
\bibitem [{\citenamefont {Kippenberg}\ \emph {et~al.}(2011)\citenamefont
  {Kippenberg}, \citenamefont {Holzwarth},\ and\ \citenamefont
  {Diddams}}]{kippenbergMicroresonatorBasedOpticalFrequency2011}%
  \BibitemOpen
  \bibfield  {author} {\bibinfo {author} {\bibfnamefont {T.~J.}\ \bibnamefont
  {Kippenberg}}, \bibinfo {author} {\bibfnamefont {R.}~\bibnamefont
  {Holzwarth}}, \ and\ \bibinfo {author} {\bibfnamefont {S.~A.}\ \bibnamefont
  {Diddams}},\ }\href {\doibase 10.1126/science.1193968} {\bibfield  {journal}
  {\bibinfo  {journal} {science}\ }\textbf {\bibinfo {volume} {332}},\ \bibinfo
  {pages} {555} (\bibinfo {year} {2011})}\BibitemShut {NoStop}%
\bibitem [{\citenamefont {Stern}\ \emph {et~al.}(2018)\citenamefont {Stern},
  \citenamefont {Ji}, \citenamefont {Okawachi}, \citenamefont {Gaeta},\ and\
  \citenamefont {Lipson}}]{sternBatteryoperatedIntegratedFrequency2018}%
  \BibitemOpen
  \bibfield  {author} {\bibinfo {author} {\bibfnamefont {B.}~\bibnamefont
  {Stern}}, \bibinfo {author} {\bibfnamefont {X.}~\bibnamefont {Ji}}, \bibinfo
  {author} {\bibfnamefont {Y.}~\bibnamefont {Okawachi}}, \bibinfo {author}
  {\bibfnamefont {A.~L.}\ \bibnamefont {Gaeta}}, \ and\ \bibinfo {author}
  {\bibfnamefont {M.}~\bibnamefont {Lipson}},\ }\href@noop {} {\bibfield
  {journal} {\bibinfo  {journal} {Nature}\ }\textbf {\bibinfo {volume} {562}},\
  \bibinfo {pages} {401} (\bibinfo {year} {2018})}\BibitemShut {NoStop}%
\bibitem [{\citenamefont {Coen}\ and\ \citenamefont
  {Erkintalo}(2015)}]{coenTemporalCavitySolitons2015}%
  \BibitemOpen
  \bibfield  {author} {\bibinfo {author} {\bibfnamefont {S.}~\bibnamefont
  {Coen}}\ and\ \bibinfo {author} {\bibfnamefont {M.}~\bibnamefont
  {Erkintalo}},\ }in\ \href {\doibase 10.1002/9783527686476.ch2} {\emph
  {\bibinfo {booktitle} {Nonlinear {{Optical Cavity Dynamics}}}}},\ \bibinfo
  {editor} {edited by\ \bibinfo {editor} {\bibfnamefont {P.}~\bibnamefont
  {Grelu}}}\ (\bibinfo  {publisher} {{Wiley-VCH Verlag GmbH \& Co. KGaA}},\
  \bibinfo {address} {{Weinheim, Germany}},\ \bibinfo {year} {2015})\ pp.\
  \bibinfo {pages} {11--40}\BibitemShut {NoStop}%
\bibitem [{\citenamefont {Garbin}\ \emph {et~al.}(2021)\citenamefont {Garbin},
  \citenamefont {Fatome}, \citenamefont {Oppo}, \citenamefont {Erkintalo},
  \citenamefont {Murdoch},\ and\ \citenamefont
  {Coen}}]{garbinDissipativePolarizationDomain2021}%
  \BibitemOpen
  \bibfield  {author} {\bibinfo {author} {\bibfnamefont {B.}~\bibnamefont
  {Garbin}}, \bibinfo {author} {\bibfnamefont {J.}~\bibnamefont {Fatome}},
  \bibinfo {author} {\bibfnamefont {G.-L.}\ \bibnamefont {Oppo}}, \bibinfo
  {author} {\bibfnamefont {M.}~\bibnamefont {Erkintalo}}, \bibinfo {author}
  {\bibfnamefont {S.~G.}\ \bibnamefont {Murdoch}}, \ and\ \bibinfo {author}
  {\bibfnamefont {S.}~\bibnamefont {Coen}},\ }\href@noop {} {\bibfield
  {journal} {\bibinfo  {journal} {Physical Review Letters}\ }\textbf {\bibinfo
  {volume} {126}},\ \bibinfo {pages} {023904} (\bibinfo {year}
  {2021})}\BibitemShut {NoStop}%
\bibitem [{\citenamefont {Yu}\ \emph {et~al.}(2021)\citenamefont {Yu},
  \citenamefont {Cole}, \citenamefont {Jung}, \citenamefont {Moille},
  \citenamefont {Srinivasan},\ and\ \citenamefont
  {Papp}}]{yuSpontaneousPulseFormation2021}%
  \BibitemOpen
  \bibfield  {author} {\bibinfo {author} {\bibfnamefont {S.-P.}\ \bibnamefont
  {Yu}}, \bibinfo {author} {\bibfnamefont {D.~C.}\ \bibnamefont {Cole}},
  \bibinfo {author} {\bibfnamefont {H.}~\bibnamefont {Jung}}, \bibinfo {author}
  {\bibfnamefont {G.~T.}\ \bibnamefont {Moille}}, \bibinfo {author}
  {\bibfnamefont {K.}~\bibnamefont {Srinivasan}}, \ and\ \bibinfo {author}
  {\bibfnamefont {S.~B.}\ \bibnamefont {Papp}},\ }\href@noop {} {\bibfield
  {journal} {\bibinfo  {journal} {Nature Photonics}\ }\textbf {\bibinfo
  {volume} {15}},\ \bibinfo {pages} {461} (\bibinfo {year} {2021})}\BibitemShut
  {NoStop}%
\bibitem [{\citenamefont {Del~Bino}\ \emph {et~al.}(2017)\citenamefont
  {Del~Bino}, \citenamefont {Silver}, \citenamefont {Stebbings},\ and\
  \citenamefont {Del'Haye}}]{delbinoSymmetryBreakingCounterpropagating2017}%
  \BibitemOpen
  \bibfield  {author} {\bibinfo {author} {\bibfnamefont {L.}~\bibnamefont
  {Del~Bino}}, \bibinfo {author} {\bibfnamefont {J.~M.}\ \bibnamefont
  {Silver}}, \bibinfo {author} {\bibfnamefont {S.~L.}\ \bibnamefont
  {Stebbings}}, \ and\ \bibinfo {author} {\bibfnamefont {P.}~\bibnamefont
  {Del'Haye}},\ }\href@noop {} {\bibfield  {journal} {\bibinfo  {journal}
  {Scientific Reports}\ }\textbf {\bibinfo {volume} {7}},\ \bibinfo {pages} {1}
  (\bibinfo {year} {2017})}\BibitemShut {NoStop}%
\bibitem [{\citenamefont {Cao}\ \emph {et~al.}(2017)\citenamefont {Cao},
  \citenamefont {Wang}, \citenamefont {Dong}, \citenamefont {Jing},
  \citenamefont {Liu}, \citenamefont {Chen}, \citenamefont {Ge}, \citenamefont
  {Gong},\ and\ \citenamefont
  {Xiao}}]{caoExperimentalDemonstrationSpontaneous2017}%
  \BibitemOpen
  \bibfield  {author} {\bibinfo {author} {\bibfnamefont {Q.-T.}\ \bibnamefont
  {Cao}}, \bibinfo {author} {\bibfnamefont {H.}~\bibnamefont {Wang}}, \bibinfo
  {author} {\bibfnamefont {C.-H.}\ \bibnamefont {Dong}}, \bibinfo {author}
  {\bibfnamefont {H.}~\bibnamefont {Jing}}, \bibinfo {author} {\bibfnamefont
  {R.-S.}\ \bibnamefont {Liu}}, \bibinfo {author} {\bibfnamefont
  {X.}~\bibnamefont {Chen}}, \bibinfo {author} {\bibfnamefont {L.}~\bibnamefont
  {Ge}}, \bibinfo {author} {\bibfnamefont {Q.}~\bibnamefont {Gong}}, \ and\
  \bibinfo {author} {\bibfnamefont {Y.-F.}\ \bibnamefont {Xiao}},\ }\href@noop
  {} {\bibfield  {journal} {\bibinfo  {journal} {Physical Review Letters}\
  }\textbf {\bibinfo {volume} {118}},\ \bibinfo {pages} {033901} (\bibinfo
  {year} {2017})},\ \Eprint {http://arxiv.org/abs/1607.01459}
  {arxiv:1607.01459} \BibitemShut {NoStop}%
\bibitem [{\citenamefont {Woodley}\ \emph {et~al.}(2018)\citenamefont
  {Woodley}, \citenamefont {Silver}, \citenamefont {Hill}, \citenamefont
  {Copie}, \citenamefont {Del~Bino}, \citenamefont {Zhang}, \citenamefont
  {Oppo},\ and\ \citenamefont
  {Del'Haye}}]{woodleyUniversalSymmetrybreakingDynamics2018}%
  \BibitemOpen
  \bibfield  {author} {\bibinfo {author} {\bibfnamefont {M.~T.}\ \bibnamefont
  {Woodley}}, \bibinfo {author} {\bibfnamefont {J.~M.}\ \bibnamefont {Silver}},
  \bibinfo {author} {\bibfnamefont {L.}~\bibnamefont {Hill}}, \bibinfo {author}
  {\bibfnamefont {F.}~\bibnamefont {Copie}}, \bibinfo {author} {\bibfnamefont
  {L.}~\bibnamefont {Del~Bino}}, \bibinfo {author} {\bibfnamefont
  {S.}~\bibnamefont {Zhang}}, \bibinfo {author} {\bibfnamefont {G.-L.}\
  \bibnamefont {Oppo}}, \ and\ \bibinfo {author} {\bibfnamefont
  {P.}~\bibnamefont {Del'Haye}},\ }\href@noop {} {\bibfield  {journal}
  {\bibinfo  {journal} {Physical Review A}\ }\textbf {\bibinfo {volume} {98}},\
  \bibinfo {pages} {053863} (\bibinfo {year} {2018})}\BibitemShut {NoStop}%
\bibitem [{\citenamefont {Garbin}\ \emph {et~al.}(2020)\citenamefont {Garbin},
  \citenamefont {Fatome}, \citenamefont {Oppo}, \citenamefont {Erkintalo},
  \citenamefont {Murdoch},\ and\ \citenamefont
  {Coen}}]{garbinAsymmetricBalanceSymmetry2020}%
  \BibitemOpen
  \bibfield  {author} {\bibinfo {author} {\bibfnamefont {B.}~\bibnamefont
  {Garbin}}, \bibinfo {author} {\bibfnamefont {J.}~\bibnamefont {Fatome}},
  \bibinfo {author} {\bibfnamefont {G.-L.}\ \bibnamefont {Oppo}}, \bibinfo
  {author} {\bibfnamefont {M.}~\bibnamefont {Erkintalo}}, \bibinfo {author}
  {\bibfnamefont {S.~G.}\ \bibnamefont {Murdoch}}, \ and\ \bibinfo {author}
  {\bibfnamefont {S.}~\bibnamefont {Coen}},\ }\href@noop {} {\bibfield
  {journal} {\bibinfo  {journal} {Physical Review Research}\ }\textbf {\bibinfo
  {volume} {2}},\ \bibinfo {pages} {023244} (\bibinfo {year}
  {2020})}\BibitemShut {NoStop}%
\bibitem [{\citenamefont {Xu}\ \emph {et~al.}(2021)\citenamefont {Xu},
  \citenamefont {Nielsen}, \citenamefont {Garbin}, \citenamefont {Hill},
  \citenamefont {Oppo}, \citenamefont {Fatome}, \citenamefont {Murdoch},
  \citenamefont {Coen},\ and\ \citenamefont
  {Erkintalo}}]{xuSpontaneousSymmetryBreaking2021}%
  \BibitemOpen
  \bibfield  {author} {\bibinfo {author} {\bibfnamefont {G.}~\bibnamefont
  {Xu}}, \bibinfo {author} {\bibfnamefont {A.~U.}\ \bibnamefont {Nielsen}},
  \bibinfo {author} {\bibfnamefont {B.}~\bibnamefont {Garbin}}, \bibinfo
  {author} {\bibfnamefont {L.}~\bibnamefont {Hill}}, \bibinfo {author}
  {\bibfnamefont {G.-L.}\ \bibnamefont {Oppo}}, \bibinfo {author}
  {\bibfnamefont {J.}~\bibnamefont {Fatome}}, \bibinfo {author} {\bibfnamefont
  {S.~G.}\ \bibnamefont {Murdoch}}, \bibinfo {author} {\bibfnamefont
  {S.}~\bibnamefont {Coen}}, \ and\ \bibinfo {author} {\bibfnamefont
  {M.}~\bibnamefont {Erkintalo}},\ }\href {\doibase 10.1038/s41467-021-24251-0}
  {\bibfield  {journal} {\bibinfo  {journal} {Nat Commun}\ }\textbf {\bibinfo
  {volume} {12}},\ \bibinfo {pages} {4023} (\bibinfo {year}
  {2021})}\BibitemShut {NoStop}%
\bibitem [{\citenamefont {{B Garbin}}\ \emph {et~al.}(2022)\citenamefont {{B
  Garbin}}, \citenamefont {{A Giraldo}}, \citenamefont {{K J H Peters}},
  \citenamefont {{N G R Broderick}}, \citenamefont {{A Spakman}}, \citenamefont
  {{F Raineri}}, \citenamefont {{A Levenson}}, \citenamefont {{S R K
  Rodriguez}}, \citenamefont {{B Krauskopf}},\ and\ \citenamefont {{A M
  Yacomotti}}}]{bgarbinSpontaneousSymmetryBreaking2022}%
  \BibitemOpen
  \bibfield  {author} {\bibinfo {author} {\bibnamefont {{B Garbin}}}, \bibinfo
  {author} {\bibnamefont {{A Giraldo}}}, \bibinfo {author} {\bibnamefont {{K J
  H Peters}}}, \bibinfo {author} {\bibnamefont {{N G R Broderick}}}, \bibinfo
  {author} {\bibnamefont {{A Spakman}}}, \bibinfo {author} {\bibnamefont {{F
  Raineri}}}, \bibinfo {author} {\bibnamefont {{A Levenson}}}, \bibinfo
  {author} {\bibnamefont {{S R K Rodriguez}}}, \bibinfo {author} {\bibnamefont
  {{B Krauskopf}}}, \ and\ \bibinfo {author} {\bibnamefont {{A M Yacomotti}}},\
  }\href {\doibase 10.1103/physrevlett.128.053901} {\bibfield  {journal}
  {\bibinfo  {journal} {Physical Review Letters}\ }\textbf {\bibinfo {volume}
  {128}},\ \bibinfo {pages} {053901} (\bibinfo {year} {2022})}\BibitemShut
  {NoStop}%
\bibitem [{\citenamefont {Haelterman}\ \emph {et~al.}(1994)\citenamefont
  {Haelterman}, \citenamefont {Trillo},\ and\ \citenamefont
  {Wabnitz}}]{haeltermanPolarizationMultistabilityInstability1994}%
  \BibitemOpen
  \bibfield  {author} {\bibinfo {author} {\bibfnamefont {M.}~\bibnamefont
  {Haelterman}}, \bibinfo {author} {\bibfnamefont {S.}~\bibnamefont {Trillo}},
  \ and\ \bibinfo {author} {\bibfnamefont {S.}~\bibnamefont {Wabnitz}},\ }\href
  {\doibase 10.1364/JOSAB.11.000446} {\bibfield  {journal} {\bibinfo  {journal}
  {J. Opt. Soc. Am. B}\ }\textbf {\bibinfo {volume} {11}},\ \bibinfo {pages}
  {446} (\bibinfo {year} {1994})}\BibitemShut {NoStop}%
\bibitem [{\citenamefont {Stojanovski}\ and\ \citenamefont
  {Kocarev}(2001)}]{stojanovskiChaosbasedRandomNumber2001}%
  \BibitemOpen
  \bibfield  {author} {\bibinfo {author} {\bibfnamefont {T.}~\bibnamefont
  {Stojanovski}}\ and\ \bibinfo {author} {\bibfnamefont {L.}~\bibnamefont
  {Kocarev}},\ }\href@noop {} {\bibfield  {journal} {\bibinfo  {journal} {IEEE
  Transactions on Circuits and Systems I: Fundamental Theory and Applications}\
  }\textbf {\bibinfo {volume} {48}},\ \bibinfo {pages} {281} (\bibinfo {year}
  {2001})}\BibitemShut {NoStop}%
\bibitem [{\citenamefont {Yuan}\ \emph {et~al.}(2015)\citenamefont {Yuan},
  \citenamefont {Cao},\ and\ \citenamefont
  {Ma}}]{yuanRandomnessRequirementClauserHorneShimonyHolt2015}%
  \BibitemOpen
  \bibfield  {author} {\bibinfo {author} {\bibfnamefont {X.}~\bibnamefont
  {Yuan}}, \bibinfo {author} {\bibfnamefont {Z.}~\bibnamefont {Cao}}, \ and\
  \bibinfo {author} {\bibfnamefont {X.}~\bibnamefont {Ma}},\ }\href@noop {}
  {\bibfield  {journal} {\bibinfo  {journal} {Physical Review A}\ }\textbf
  {\bibinfo {volume} {91}},\ \bibinfo {pages} {032111} (\bibinfo {year}
  {2015})}\BibitemShut {NoStop}%
\bibitem [{\citenamefont {Crauel}\ and\ \citenamefont
  {Flandoli}(1998)}]{crauelAdditiveNoiseDestroys1998}%
  \BibitemOpen
  \bibfield  {author} {\bibinfo {author} {\bibfnamefont {H.}~\bibnamefont
  {Crauel}}\ and\ \bibinfo {author} {\bibfnamefont {F.}~\bibnamefont
  {Flandoli}},\ }\href {\doibase 10.1023/A:1022665916629} {\bibfield  {journal}
  {\bibinfo  {journal} {Journal of Dynamics and Differential Equations}\
  }\textbf {\bibinfo {volume} {10}},\ \bibinfo {pages} {259} (\bibinfo {year}
  {1998})}\BibitemShut {NoStop}%
\bibitem [{\citenamefont {Coen}\ \emph {et~al.}(2023)\citenamefont {Coen},
  \citenamefont {Garbin}, \citenamefont {Xu}, \citenamefont {Quinn},
  \citenamefont {Goldman}, \citenamefont {Oppo}, \citenamefont {Erkintalo},
  \citenamefont {Murdoch},\ and\ \citenamefont
  {Fatome}}]{coenNonlinearTopologicalSymmetry2023}%
  \BibitemOpen
  \bibfield  {author} {\bibinfo {author} {\bibfnamefont {S.}~\bibnamefont
  {Coen}}, \bibinfo {author} {\bibfnamefont {B.}~\bibnamefont {Garbin}},
  \bibinfo {author} {\bibfnamefont {G.}~\bibnamefont {Xu}}, \bibinfo {author}
  {\bibfnamefont {L.}~\bibnamefont {Quinn}}, \bibinfo {author} {\bibfnamefont
  {N.}~\bibnamefont {Goldman}}, \bibinfo {author} {\bibfnamefont {G.-L.}\
  \bibnamefont {Oppo}}, \bibinfo {author} {\bibfnamefont {M.}~\bibnamefont
  {Erkintalo}}, \bibinfo {author} {\bibfnamefont {S.~G.}\ \bibnamefont
  {Murdoch}}, \ and\ \bibinfo {author} {\bibfnamefont {J.}~\bibnamefont
  {Fatome}},\ }\href@noop {} {\bibfield  {journal} {\bibinfo  {journal} {arXiv
  preprint arXiv:2303.16197}\ } (\bibinfo {year} {2023})},\ \Eprint
  {http://arxiv.org/abs/2303.16197} {arxiv:2303.16197} \BibitemShut {NoStop}%
\bibitem [{\citenamefont {Steinmeyer}\ \emph {et~al.}(1994)\citenamefont
  {Steinmeyer}, \citenamefont {Jaspert},\ and\ \citenamefont
  {Mitschke}}]{steinmeyerObservationPerioddoublingSequence1994}%
  \BibitemOpen
  \bibfield  {author} {\bibinfo {author} {\bibfnamefont {G.}~\bibnamefont
  {Steinmeyer}}, \bibinfo {author} {\bibfnamefont {D.}~\bibnamefont {Jaspert}},
  \ and\ \bibinfo {author} {\bibfnamefont {F.}~\bibnamefont {Mitschke}},\
  }\href {\doibase 10.1016/0030-4018(94)90574-6} {\bibfield  {journal}
  {\bibinfo  {journal} {Optics Communications}\ }\textbf {\bibinfo {volume}
  {104}},\ \bibinfo {pages} {379} (\bibinfo {year} {1994})}\BibitemShut
  {NoStop}%
\bibitem [{\citenamefont {Steinle}\ \emph {et~al.}(2017)\citenamefont
  {Steinle}, \citenamefont {Greiner}, \citenamefont {Wrachtrup}, \citenamefont
  {Giessen},\ and\ \citenamefont
  {Gerhardt}}]{steinleUnbiasedAllOpticalRandomNumber2017}%
  \BibitemOpen
  \bibfield  {author} {\bibinfo {author} {\bibfnamefont {T.}~\bibnamefont
  {Steinle}}, \bibinfo {author} {\bibfnamefont {J.~N.}\ \bibnamefont
  {Greiner}}, \bibinfo {author} {\bibfnamefont {J.}~\bibnamefont {Wrachtrup}},
  \bibinfo {author} {\bibfnamefont {H.}~\bibnamefont {Giessen}}, \ and\
  \bibinfo {author} {\bibfnamefont {I.}~\bibnamefont {Gerhardt}},\ }\href
  {\doibase 10.1103/PhysRevX.7.041050} {\bibfield  {journal} {\bibinfo
  {journal} {Phys. Rev. X}\ }\textbf {\bibinfo {volume} {7}},\ \bibinfo {pages}
  {041050} (\bibinfo {year} {2017})}\BibitemShut {NoStop}%
\bibitem [{\citenamefont {Rukhin}\ \emph {et~al.}(2001)\citenamefont {Rukhin},
  \citenamefont {Soto}, \citenamefont {Nechvatal}, \citenamefont {Smid},\ and\
  \citenamefont {Barker}}]{rukhinStatisticalTestSuite2001}%
  \BibitemOpen
  \bibfield  {author} {\bibinfo {author} {\bibfnamefont {A.}~\bibnamefont
  {Rukhin}}, \bibinfo {author} {\bibfnamefont {J.}~\bibnamefont {Soto}},
  \bibinfo {author} {\bibfnamefont {J.}~\bibnamefont {Nechvatal}}, \bibinfo
  {author} {\bibfnamefont {M.}~\bibnamefont {Smid}}, \ and\ \bibinfo {author}
  {\bibfnamefont {E.}~\bibnamefont {Barker}},\ }\href@noop {} {\emph {\bibinfo
  {title} {A Statistical Test Suite for Random and Pseudorandom Number
  Generators for Cryptographic Applications}}},\ \bibinfo {type} {Tech. Rep.}\
  (\bibinfo  {institution} {{National Institute of Standards and Technology}},\
  \bibinfo {year} {2001})\BibitemShut {NoStop}%
\bibitem [{\citenamefont {Brown}\ \emph {et~al.}(2018)\citenamefont {Brown},
  \citenamefont {Eddelbuettel},\ and\ \citenamefont
  {Bauer}}]{brownDieharder2018}%
  \BibitemOpen
  \bibfield  {author} {\bibinfo {author} {\bibfnamefont {R.~G.}\ \bibnamefont
  {Brown}}, \bibinfo {author} {\bibfnamefont {D.}~\bibnamefont {Eddelbuettel}},
  \ and\ \bibinfo {author} {\bibfnamefont {D.}~\bibnamefont {Bauer}},\
  }\href@noop {} {\bibfield  {journal} {\bibinfo  {journal} {Duke University
  Physics Department}\ ,\ \bibinfo {pages} {27708}} (\bibinfo {year}
  {2018})}\BibitemShut {NoStop}%
\bibitem [{\citenamefont {Marandi}\ \emph {et~al.}(2012)\citenamefont
  {Marandi}, \citenamefont {Leindecker}, \citenamefont {Vodopyanov},\ and\
  \citenamefont {Byer}}]{marandiAllopticalQuantumRandom2012}%
  \BibitemOpen
  \bibfield  {author} {\bibinfo {author} {\bibfnamefont {A.}~\bibnamefont
  {Marandi}}, \bibinfo {author} {\bibfnamefont {N.~C.}\ \bibnamefont
  {Leindecker}}, \bibinfo {author} {\bibfnamefont {K.~L.}\ \bibnamefont
  {Vodopyanov}}, \ and\ \bibinfo {author} {\bibfnamefont {R.~L.}\ \bibnamefont
  {Byer}},\ }\href {\doibase 10.1364/OE.20.019322} {\bibfield  {journal}
  {\bibinfo  {journal} {Opt. Express}\ }\textbf {\bibinfo {volume} {20}},\
  \bibinfo {pages} {19322} (\bibinfo {year} {2012})}\BibitemShut {NoStop}%
\bibitem [{\citenamefont {Okawachi}\ \emph {et~al.}(2016)\citenamefont
  {Okawachi}, \citenamefont {Yu}, \citenamefont {Luke}, \citenamefont
  {Carvalho}, \citenamefont {Lipson},\ and\ \citenamefont
  {Gaeta}}]{okawachiQuantumRandomNumber2016}%
  \BibitemOpen
  \bibfield  {author} {\bibinfo {author} {\bibfnamefont {Y.}~\bibnamefont
  {Okawachi}}, \bibinfo {author} {\bibfnamefont {M.}~\bibnamefont {Yu}},
  \bibinfo {author} {\bibfnamefont {K.}~\bibnamefont {Luke}}, \bibinfo {author}
  {\bibfnamefont {D.~O.}\ \bibnamefont {Carvalho}}, \bibinfo {author}
  {\bibfnamefont {M.}~\bibnamefont {Lipson}}, \ and\ \bibinfo {author}
  {\bibfnamefont {A.~L.}\ \bibnamefont {Gaeta}},\ }\href {\doibase
  10.1364/OL.41.004194} {\bibfield  {journal} {\bibinfo  {journal} {Opt.
  Lett.}\ }\textbf {\bibinfo {volume} {41}},\ \bibinfo {pages} {4194} (\bibinfo
  {year} {2016})}\BibitemShut {NoStop}%
\bibitem [{\citenamefont {Wang}\ \emph {et~al.}(2013)\citenamefont {Wang},
  \citenamefont {Marandi}, \citenamefont {Wen}, \citenamefont {Byer},\ and\
  \citenamefont {Yamamoto}}]{wangCoherentIsingMachine2013}%
  \BibitemOpen
  \bibfield  {author} {\bibinfo {author} {\bibfnamefont {Z.}~\bibnamefont
  {Wang}}, \bibinfo {author} {\bibfnamefont {A.}~\bibnamefont {Marandi}},
  \bibinfo {author} {\bibfnamefont {K.}~\bibnamefont {Wen}}, \bibinfo {author}
  {\bibfnamefont {R.~L.}\ \bibnamefont {Byer}}, \ and\ \bibinfo {author}
  {\bibfnamefont {Y.}~\bibnamefont {Yamamoto}},\ }\href {\doibase
  10.1103/PhysRevA.88.063853} {\bibfield  {journal} {\bibinfo  {journal} {Phys.
  Rev. A}\ }\textbf {\bibinfo {volume} {88}},\ \bibinfo {pages} {063853}
  (\bibinfo {year} {2013})}\BibitemShut {NoStop}%
\bibitem [{\citenamefont {Inagaki}\ \emph {et~al.}(2016)\citenamefont
  {Inagaki}, \citenamefont {Haribara}, \citenamefont {Igarashi}, \citenamefont
  {Sonobe}, \citenamefont {Tamate}, \citenamefont {Honjo}, \citenamefont
  {Marandi}, \citenamefont {McMahon}, \citenamefont {Umeki},\ and\
  \citenamefont {Enbutsu}}]{inagakiCoherentIsingMachine2016}%
  \BibitemOpen
  \bibfield  {author} {\bibinfo {author} {\bibfnamefont {T.}~\bibnamefont
  {Inagaki}}, \bibinfo {author} {\bibfnamefont {Y.}~\bibnamefont {Haribara}},
  \bibinfo {author} {\bibfnamefont {K.}~\bibnamefont {Igarashi}}, \bibinfo
  {author} {\bibfnamefont {T.}~\bibnamefont {Sonobe}}, \bibinfo {author}
  {\bibfnamefont {S.}~\bibnamefont {Tamate}}, \bibinfo {author} {\bibfnamefont
  {T.}~\bibnamefont {Honjo}}, \bibinfo {author} {\bibfnamefont
  {A.}~\bibnamefont {Marandi}}, \bibinfo {author} {\bibfnamefont {P.~L.}\
  \bibnamefont {McMahon}}, \bibinfo {author} {\bibfnamefont {T.}~\bibnamefont
  {Umeki}}, \ and\ \bibinfo {author} {\bibfnamefont {K.}~\bibnamefont
  {Enbutsu}},\ }\href@noop {} {\bibfield  {journal} {\bibinfo  {journal}
  {Science}\ }\textbf {\bibinfo {volume} {354}},\ \bibinfo {pages} {603}
  (\bibinfo {year} {2016})}\BibitemShut {NoStop}%
\bibitem [{\citenamefont {McMahon}\ \emph {et~al.}(2016)\citenamefont
  {McMahon}, \citenamefont {Marandi}, \citenamefont {Haribara}, \citenamefont
  {Hamerly}, \citenamefont {Langrock}, \citenamefont {Tamate}, \citenamefont
  {Inagaki}, \citenamefont {Takesue}, \citenamefont {Utsunomiya}, \citenamefont
  {Aihara}, \citenamefont {Byer}, \citenamefont {Fejer}, \citenamefont
  {Mabuchi},\ and\ \citenamefont
  {Yamamoto}}]{mcmahonFullyProgrammable100spin2016}%
  \BibitemOpen
  \bibfield  {author} {\bibinfo {author} {\bibfnamefont {P.~L.}\ \bibnamefont
  {McMahon}}, \bibinfo {author} {\bibfnamefont {A.}~\bibnamefont {Marandi}},
  \bibinfo {author} {\bibfnamefont {Y.}~\bibnamefont {Haribara}}, \bibinfo
  {author} {\bibfnamefont {R.}~\bibnamefont {Hamerly}}, \bibinfo {author}
  {\bibfnamefont {C.}~\bibnamefont {Langrock}}, \bibinfo {author}
  {\bibfnamefont {S.}~\bibnamefont {Tamate}}, \bibinfo {author} {\bibfnamefont
  {T.}~\bibnamefont {Inagaki}}, \bibinfo {author} {\bibfnamefont
  {H.}~\bibnamefont {Takesue}}, \bibinfo {author} {\bibfnamefont
  {S.}~\bibnamefont {Utsunomiya}}, \bibinfo {author} {\bibfnamefont
  {K.}~\bibnamefont {Aihara}}, \bibinfo {author} {\bibfnamefont {R.~L.}\
  \bibnamefont {Byer}}, \bibinfo {author} {\bibfnamefont {M.~M.}\ \bibnamefont
  {Fejer}}, \bibinfo {author} {\bibfnamefont {H.}~\bibnamefont {Mabuchi}}, \
  and\ \bibinfo {author} {\bibfnamefont {Y.}~\bibnamefont {Yamamoto}},\ }\href
  {\doibase 10.1126/science.aah5178} {\bibfield  {journal} {\bibinfo  {journal}
  {Science}\ }\textbf {\bibinfo {volume} {354}},\ \bibinfo {pages} {614}
  (\bibinfo {year} {2016})}\BibitemShut {NoStop}%
\bibitem [{\citenamefont {{Honari-Latifpour}}\ and\ \citenamefont
  {Miri}(2020)}]{honari-latifpourOpticalPottsMachine2020}%
  \BibitemOpen
  \bibfield  {author} {\bibinfo {author} {\bibfnamefont {M.}~\bibnamefont
  {{Honari-Latifpour}}}\ and\ \bibinfo {author} {\bibfnamefont {M.-A.}\
  \bibnamefont {Miri}},\ }\href@noop {} {\bibfield  {journal} {\bibinfo
  {journal} {Nanophotonics}\ }\textbf {\bibinfo {volume} {9}},\ \bibinfo
  {pages} {4199} (\bibinfo {year} {2020})}\BibitemShut {NoStop}%
\bibitem [{\citenamefont {Lugiato}\ and\ \citenamefont
  {Lefever}(1987)}]{lugiatoSpatialDissipativeStructures1987}%
  \BibitemOpen
  \bibfield  {author} {\bibinfo {author} {\bibfnamefont {L.~A.}\ \bibnamefont
  {Lugiato}}\ and\ \bibinfo {author} {\bibfnamefont {R.}~\bibnamefont
  {Lefever}},\ }\href@noop {} {\bibfield  {journal} {\bibinfo  {journal}
  {Physical review letters}\ }\textbf {\bibinfo {volume} {58}},\ \bibinfo
  {pages} {2209} (\bibinfo {year} {1987})}\BibitemShut {NoStop}%
\bibitem [{\citenamefont {Nielsen}\ \emph {et~al.}(2018)\citenamefont
  {Nielsen}, \citenamefont {Garbin}, \citenamefont {Coen}, \citenamefont
  {Murdoch},\ and\ \citenamefont
  {Erkintalo}}]{nielsenInvitedArticleEmission2018}%
  \BibitemOpen
  \bibfield  {author} {\bibinfo {author} {\bibfnamefont {A.~U.}\ \bibnamefont
  {Nielsen}}, \bibinfo {author} {\bibfnamefont {B.}~\bibnamefont {Garbin}},
  \bibinfo {author} {\bibfnamefont {S.}~\bibnamefont {Coen}}, \bibinfo {author}
  {\bibfnamefont {S.~G.}\ \bibnamefont {Murdoch}}, \ and\ \bibinfo {author}
  {\bibfnamefont {M.}~\bibnamefont {Erkintalo}},\ }\href {\doibase
  10.1063/1.5060123} {\bibfield  {journal} {\bibinfo  {journal} {APL
  Photonics}\ }\textbf {\bibinfo {volume} {3}},\ \bibinfo {pages} {120804}
  (\bibinfo {year} {2018})}\BibitemShut {NoStop}%
\end{thebibliography}%

% Full bibliography added automatically for Optics Letters submissions; the following line will simply be ignored if submitting to other journals.
% Note that this extra page will not count against page length
%\bibliographyfullrefs{bibliography_randomness}

\end{document}